\documentstyle[12pt]{article}

\def\s{{\rm \;sn}}
\def\dd{{\rm \;dn}}
\def\cc{{\rm \;cn}}
\def\bp{{\rm b}^{+}}
\def\bm{{\rm b}^{-}}
\def\cp{{\rm c}^{+}}
\def\cm{{\rm c}^{-}}
\def\dpl{{\rm d}^{+}}
\def\dm{{\rm d}^{-}}
\def\b{{\rm b}^{\pm}}
\def\c{{\rm c}^{\pm}}
\def\d{{\rm d}^{\pm}}
\def\eu{\epsilon_{1}}
\def\ed{\epsilon_{2}}
\def\et{\epsilon_{3}}

\begin{document}

\begin{titlepage}
\title{Boundary K-matrices for the XYZ, XXZ and XXX spin chains}
\author{H.J. de Vega\\
A. Gonz\'alez--Ruiz\thanks{Permanent address: Departamento
de F\'{\i}sica Te\'orica, Universidad Complutense, 28040 Madrid,
ESPA\~NA} \\
{\it L.P.T.H.E.} \\
{\it Tour 16, 1er \'etage, Universit\'e Paris VI} \\
{\it 4 Place Jussieu, 75252 Paris cedex 05, FRANCE}}
\date{}
\maketitle

\begin {abstract}
The general solutions for the factorization equations
of the reflection matrices $K^{\pm}(\theta)$ for the eight vertex and
six vertex models
 (XYZ, XXZ and XXX chains)
are found. The associated integrable magnetic Hamiltonians
are explicitly derived, finding families dependig on several continuous
as well as discrete  parameters.
\end{abstract}

\vskip-16.0cm
\rightline{}
\rightline{{\bf LPTHE--PAR 93/29}}
\rightline{{\bf June 1993}}
\vskip2cm

\end{titlepage}

\begin{section}{Introduction}
It is clearly interesting to find the widest possible class of
boundary conditions
compatible with integrability asocciated to a given model.\\
Not any boundary condition (b.c.) obeys this requirement.
Periodic  and twisted (under a symmetry of the model) b.c. are usually
compatible with the Yang-Baxter equations \cite{B,h}. In addition, there
are the b.c. defined by reflection matrices $K^{\pm}$
\cite{sk,nos,mn,fr}. These  $K^{\pm}$ matrices can be interpreted as
defining the scattering by the boundaries. In a recent publication
\cite{za} the interpretation of this matrices as boundary
S-matrices in two dimensional integrable quantum
 field theories was developed. They also imply boundary
terms for the spin hamiltonians which can be interpreted as the
coupling with magnetic
fields in the edges of the chain.\\
In addition, quantum group invariance arises
for specific choices of fixed b.c. (See for example \cite{ps,qba1,mn} for
the trigonometric case and \cite{royo} for the elliptic one). A
quantum group--like structure is still is to be found for which
 Baxter's 8-vertex elliptic matrix
\cite{B} could act as an intertwiner
(for a recent attempt see \cite{U}) giving an affine quantum
invariance to the infinite spin chain and the boundary terms for
the quantum group invariance of the finite chain. This program has
been done in the elliptic case for the free fermionic model, see
\cite{rogo,royo}.\\
A general setting to find boundary terms compatible with
integrability was proposed by Sklyanin \cite{sk} .
To find these boundary conditions one has to solve the so called reflection
equations:

\begin{equation}
\begin{array}{l}
R(\theta-\theta^\prime)[K^{-}(\theta)\otimes
1]R(\theta+\theta^\prime)[K^{-}(\theta^\prime)\otimes 1]=\\
{[}K^{-}(\theta^\prime)\otimes 1]R(\theta+\theta^\prime)[K^{-}(\theta)\otimes
1]R(\theta-\theta^\prime)
\;\;, \label{sk1}
\end{array}
\end{equation}

\begin{equation}
\begin{array}{l}
R(\theta-\theta^\prime)[1\otimes K^{+}(\theta) ]R(\theta+\theta^\prime)[
1\otimes K^{+}(\theta^\prime)]=\\ {[} 1\otimes K^{+}(\theta^\prime)]
R(\theta+\theta^\prime)[1\otimes K^{+}(\theta) ]R(\theta-\theta^\prime)
\;\;,\label{sk2}
\end{array}
\end{equation}

where $R(\theta)$ is the  R-matrix of the chain and $K^{\pm}(\theta)$
give the boundary terms (see below).

As is known, the XYZ model is obtained from the elliptic eight-vertex
 solution of the Yang-Baxter equation:

\begin{equation}
\begin{array}{l}
[1\otimes R(\theta-\theta^\prime)][R(\theta)\otimes 1][1\otimes
R(\theta^\prime)]=\\
{[}R(\theta^\prime)\otimes 1][1\otimes R(\theta)]
[R(\theta-\theta^\prime)\otimes 1]
\;\;.\label{yb}
\end{array}
\end{equation}

The XXZ and XXX models follow respectively from the
trigonometric and rational limits of this R-matrix.\\
We present in this paper the general solutions $K^{\pm}(\theta)$ to these
equations for the XYZ, XXZ and XXX models. We find for the elliptic
case two families of solutions, each family depending
 on one continuous and
one discrete parameter, see equations
(\ref{sola}) and (\ref{solb}). For the trigonometric
 and rational limit we
find a family of solutions depending  on four continuous parameters,
see equations (\ref{KG}) and
(\ref{Kr}) respectively.\\
We remark that the trigonometric limit of the elliptic
solutions of  (\ref{sk1}),(\ref{sk2}) does not provide all solutions
to the trigonometric/ hyperbolic case.\\
{F}rom these $K^{\pm}(\theta)$ solutions we derive the boundary terms
in the XYZ
hamiltonian wich are compatible with integrability. Finally we analyze
the relation of the present eight vertex results with the general
K-matrices of the six-vertex reported in ref. \cite{nos} and consider
in addition the rational limit.

\end{section}

\begin{section}{General solution to the reflection equations for the
Eight Vertex model (XYZ chain)}
The R-matrix for the XYZ chain can be written as \cite{B}:

\begin{equation}
R(\theta)=\left(\begin{array}{cccc}
1 & 0 & 0 & k \s\gamma \s\theta\\
0 & \frac{\s\gamma}{\s(\theta+\gamma)} &
\frac{\s\theta}{\s(\theta+\gamma)} & 0\\
 0 & \frac{\s\theta}  {\s(\theta+\gamma)} &
\frac{\s\gamma}  {\s(\theta+\gamma)}  & 0\\
 k \s\gamma \s\theta & 0 & 0 & 1
\end{array}\right)
\;\;,\label{r}
\end{equation}

where $\s$ (and $\cc,\dd$ in the formulas below) stand for Jacobi
elliptic functions of modulus $0\leq k\leq 1$.\\
This solution of the Yang-Baxter equations enjoys the following
properties\\
a) Regularity: $R(0)=1$.\\
b) Parity invariance: $PR(\theta)P=R(\theta)$ where
$P^{ab}_{cd}=\delta^{a}_{d}\delta^{b}_{c}$.\\
c) Time reversal invariance: $R^{ab}_{cd}=R^{cd}_{ab}$.\\
d) Crossing unitarity: $\hat{R}(\theta)\hat{R}(-\theta-2\eta)
=\hat{\rho}(\theta)\;1$.\\
Where $\hat{R}_{cd}^{ab}=R^{ac}_{bd}$, $\eta=\gamma$ and:

\begin{eqnarray}
\hat{\rho}(\theta)&=&1-\frac{\s^{2}\gamma}{\s^{2}(\gamma+\theta)}\nonumber\\
&=&\frac{\s(\theta+2\gamma)\s\theta}
{\s^{2}(\gamma+\theta)}[1-k^{2}\s^{2}\gamma\s^{2}(\gamma+\theta)]
\;\;.
\end{eqnarray}

{F}rom (\ref{yb}) and a) unitarity follows :

\begin{eqnarray}
R(\theta)R(-\theta)&=&\rho(\theta)1\label{unit}\\
\rho(\theta)&=&1-k^{2}\s^{2}\gamma\s^{2}\theta\nonumber\\
&=&\frac{\s^{2}\gamma-\s^2\theta}{\s(\gamma+\theta)\s(\gamma-\theta)}\;\;\;.
\end{eqnarray}

It is shown in \cite{sk} that when the R-matrix enjoys properties
b),c),d) and (\ref{unit}) we can look for solutions to equations
(\ref{sk1}) and (\ref{sk2}) in order to find open boundary conditions
compatible with integrability .\\
Since b) holds, equations (\ref{sk1}) and  (\ref{sk2}) are
equivalent. We now look for the general solution of these equations in
the form:

\begin{equation}
K(\theta)=\left(\begin{array}{cc}
x(\theta) & y(\theta)\\
z(\theta) & v(\theta)
\end{array}\right)
\;\;.\label{k}
\end{equation}

Inserting equations (\ref{r}) and (\ref{k}) in (\ref{sk1}) we find
twelve independent equations:

\begin{eqnarray}
&&\bp yz'+\cp\dm z z'=\cp\dm y y'+\bp z y'\label{1}\\
&&\dm vv'+\dpl xv'=\dpl vx'+\dm xx'\label{2}\\
&&\bm yz'+\cm\dpl zz'=\cm\dpl yy'+\bm zy'\label{3}\\
&&\bp\cm vv'+\bm\cp xv'=\cp\bm vx'+\cm\bp xx'\label{4}\\
&&\cp yx'+\bp\dm zx'+\dm vz'+\dpl xz'=\nonumber\\
&&\;\;\;\cm yx'+\bp\cm vy'+\bm\cp
xy'+\bm\dpl zx'\label{5}\\
&&\bp yv'+\dm\dpl vy'+xy'+\cp\dm zv'=\nonumber\\
&&\;\;\;\bm yx'+\bm\bp vy'+\cm\cp
xy'+\cm\dpl zx'\label{6}\\
&&\bm\dpl yx'+\cm zx'+\bp\cm vz'+\bm\cp xz'=\nonumber\\
&&\;\;\;\bp\dm yx'+\dm vy'+\dpl
xy'+\cp zx'\label{7}\\
&&\bm yv'+\cm\cp vy'+\bm\bp xy'+\cm\dpl zv'=\nonumber\\
&&\;\;\;\bp yx'+vy'+\dm\dpl
xy'+\cp\dm zx'\label{8}\\
&&\cm\dpl yx'+\bm zx'+\bm\bp vz'+\cm\cp xz'=\nonumber\\
&&\;\;\;\cp\dm yv'+\bp
zv'+z'x+\dpl\dm vz'\label{9}\\
&&\cm yv'+\bm\cp vy'+\bp\cm xy'+\bm\dpl zv'=\nonumber\\
&&\;\;\;\cp yv'+\bp\dm zv'+\dpl
vz'+\dm xz'\label{10}\\
&&\cp\dm yx'+\bp zx'+vz'+\dm\dpl xz'=\nonumber\\
&&\;\;\;\cm\dpl yv'+\bm zv'+\cm\cp
vz'+\bm\bp xz'\label{11}\\
&&\bp\dm yv'+\dpl vy'+\dm xy'+\cp zv'=\nonumber\\
&&\;\;\;\bm\dpl yv'+\cm zv'+\cp\bm
vz'+\bp\cm xz'\label{12}
\end{eqnarray}

where:

\begin{equation}
R(\theta\pm\theta')=\left(\begin{array}{cccc}
1 & 0 & 0 & \d\\
0 & \b & \c & 0\\
0 & \c & \b & 0\\
\d & 0 & 0 & 1
\end{array}\right)
\;\;\;, \label{rpm}
\end{equation}

and $x'=x(\theta')$,  $y'=y(\theta')$, etc.\\
We start by assuming one of the elements of $K$ in equation (\ref{k})
 is equal to zero. There will be
four cases depending on which element is zero, but only two of them
turn out to be different:\\
a)$x=0=x'$\\
 Using equation (\ref{2}) we have $v=0=v'$ and we are left just
 with equations (\ref{3}) and (\ref{1}) as independent equations.
In order that these two
equations  be satisfied it must be that:

\begin{eqnarray}
z'/y'&=&\frac{\cm\dpl+\bm z/y}{\bm+\cm\dpl z/y}\nonumber\\
&=&\frac{\cp\dm+\bp z/y}{\bp+\cp\dm z/y}
\;\;\;,
\end{eqnarray}

which implies $(z/y)^{2}=1$. Two solutions are then obtained:

\begin{equation}
K(\theta)=\left(\begin{array}{cc}
0& 1\\
\pm 1& 0
\end{array}\right)
\;\;\;,\label{kc1}
\end{equation}

where from now on an arbitrary multiplicative  function of $\theta$
will be omitted.\\
The case where $v=0=v'$ is equivalent to this.\\
b)$z=0=z'$\\
{F}rom eq.(\ref{1}) $y=0=y'$ and from eqs.(\ref{4}) and (\ref{2}):

\begin{eqnarray}
v'/x'&=&\frac{\cp\bm v/x+\cm\bp}{\bp\cm v/x + \bm\cp}\nonumber\\
&=&\frac{\dpl v/x+\dm}{\dm v/x+\dpl}
\;\;\;,
\end{eqnarray}

which implies $(v/x)^{2}=1$, and then:

\begin{equation}
K(\theta)=\left(\begin{array}{cc}
1& 0\\
0& \pm 1
\end{array}\right)
\;\;\;.\label{kc2}
\end{equation}

The case where $y=0=y'$ is equivalent to this.\\
We now assume $x(\theta)\neq 0$ and $y(\theta)\neq 0$. Then equations
(\ref{3}) and (\ref{1}) imply that:

\begin{equation}
z(\theta)=\pm y(\theta)\neq 0\;\;\;,
\end{equation}

and (\ref{4}) and (\ref{2}) require:

\begin{equation}
v(\theta)=\pm x(\theta)\neq 0\;\;\;.
\end{equation}

The matrices $K(\theta)$ in this case have the form:

\begin{equation}
K(\theta)=\left(\begin{array}{cc}
x(\theta) & \eu y(\theta)\\
\ed z(\theta) & \et v(\theta)
\end{array}\right)
\;\;\;,\label{kp}
\end{equation}

where $\eu^{2},\ed^{2},\et^{2}=1$. Omitting an arbitrary
multiplicative
function of $\theta$ we have only eight different possibilities:\\
a) $\eu=\ed=\et=1$\\
b) $\eu=\ed=1$ and $\et=-1$\\
c) $\eu=\et=1$ and $\ed=-1$\\
d) $\ed=\et=1$ and $\eu=-1$\\
e) $\eu=1$ and $\ed=\et=-1$\\
f) $\ed=1$ and $\eu=\et=-1$\\
g) $\et=1$ and $\eu=\ed=-1$\\
h) $\eu=\ed=\et=-1$\\
Inserting (\ref{kp}) in the rest of the equations, one finds
 only two different equations for
$w(\theta)\equiv y(\theta)/x(\theta)$ in all cases. They  are:

\begin{eqnarray}
w(\theta)/w(\theta')&=&\frac{\et\bp\cm+\bm\cp-\eu\ed\dpl-\eu\ed\et\dm}
{\cp+\eu\ed\bp\dm-\cm-\eu\ed\bm\dpl}
\;\;\;,\label{pr1}
\end{eqnarray}

\begin{eqnarray}
w(\theta)/w(\theta')&=&\frac{\bm\bp+\et\cm\cp-\dm\dpl-\et}
{\bp+\eu\ed\cp\dm-\et\bm-\eu\ed\et\cm\dpl}
\;\;\;.\label{pr2}
\end{eqnarray}

For the previous equations to have a solution, the r.h.s. of
(\ref{pr1}) and (\ref{pr2}) must be identical. This can be seen,
with some work, to occur for all cases a)-h). One can see also that
 these expressions factorize as:

\begin{equation}
w(\theta)/w(\theta')=\frac{\s\theta}{[1+\eu\ed k\s^{2}\theta]}/
\frac{\s\theta'}{[1+\eu\ed k\s^{2}\theta']}\;\;\;,
\end{equation}

for the cases where $\et=1$, and as:

\begin{equation}
w(\theta)/w(\theta')=\frac{\cc\theta\dd\theta}
{[1+\eu\ed k\s^{2}\theta]}/
\frac{\cc\theta'\dd\theta'}
{[1+\eu\ed k\s^{2}\theta']}\;\;\;,
\end{equation}

for cases where $\et=-1$. We therefore have a $\theta$-independent
free parameter in the general solution that we call  $\lambda$. The solution
then reads:

\begin{equation}
w(\theta)=\lambda\frac{\s\theta}{[1+\eu\ed k\s^{2}\theta]}\;\;\;,
\end{equation}

when $\et=1$, and:

\begin{equation}
w(\theta)=\lambda\frac{\cc\theta\dd\theta}
{[1+\eu\ed k\s^{2}\theta]}\;\;\;,
\end{equation}

when $\et=-1$.\\
It can be noticed that these solutions are easily obtained by the residue
of (\ref{pr1}) when $\theta\rightarrow 0$ if $\et=1$, or the limit
$\theta\rightarrow 0$ of the same equation when $\et=-1$.\\
We summarize the general solution of the factorization equations
for the 8-vertex model as:

\begin{eqnarray}
K_{A}(\theta)=\left(\begin{array}{cc}
[1+\epsilon k \s^2\theta]& \epsilon\lambda_{HA}\s\theta\\
\lambda_{HA}\s\theta& [1+\epsilon k \s^2\theta]
\end{array}\right)
\;\;\;,\label{sola}
\end{eqnarray}

and:

\begin{eqnarray}
K_{B}(\theta)=\left(\begin{array}{cc}
[1+\epsilon k \s^2\theta]& \epsilon\lambda_{HB}\cc\theta\dd\theta\\
\lambda_{HB}\cc\theta\dd\theta& -[1+\epsilon k \s^2\theta]
\end{array}\right)
\;\;\;,\label{solb}
\end{eqnarray}

where $\epsilon^{2}=1$ and $\lambda_{HA},\lambda_{HB}$ are arbitrary
 parameters. That is, we find two families of solutions
 each one depending on a continuous and on a discrete parameter. (The
discrete parameter takes
only two values).\\
These solutions lead in the trigonometric limit $k=0$ to only some
 specific cases of the general solution for the six vertex
R-matrix as discussed in the next section.\\
We now look for the hamiltonians obtained by the first derivative
of the transfer matrix \cite{sk}:

\begin{equation}
H=\sum^{N-1}_{n=1}h_{n,n+1}+\frac{1}{2} (K^{-}_{1}(0)^{-1})\dot{K}^{-}_{1}(0)+
\frac{tr_{0}[K^{+t}_{0}(-\eta)h_{N0}]}{tr[K^{+}(-\eta)]}
\;\;\;,\label{H}
\end{equation}

 where:

\begin{equation}
h_{n,n+1}=\dot{R}_{n,n+1}(0)\;\;\;,
\end{equation}

and the term $(K^{-}_{1}(0)^{-1})\dot{K}^{-}_{1}(0)$ generalizes the
formula for the hamiltonian of Sklyanin to the case when $K^{-}(0)\neq 1$.
This formula is only defined when $tr[K^{+}(-\eta)]\neq 0$ and
$det[K^{-}(0)]\neq 0$.\\
 We see in equation (\ref{solb}) that for the second family of solutions the
trace of $K$ is zero. For this second family we will then not have a
 well defined
hamiltonian from the first derivative of the transfer matrix.\\
 When

\begin{equation}
tr[K^{+}(-\eta)]=0
\;\;\;,\label{trcero}
\end{equation}

and:

\begin{equation}
tr_{0}[K^{+t}_{0}(-\eta)h_{N0}]\propto 1
\;\;\;,\label{pro1}
\end{equation}

a well defined hamiltonian
with only nearest neighbours interactions is obtained from the second
derivative of the transfer matrix as shown in \cite{royo}. But for
the present second familiy of solutions the  condition (\ref{pro1})
does not hold.
Furthermore $\dot{K_{B}}(0)=0$ wich gives only a trivial boundary term at
the left end. The same happens with solutions (\ref{kc1}),(\ref{kc2})
  where one of the elements is zero. \\
If the condition (\ref{trcero}) holds but not eq.(\ref{pro1}), one obtains
from the second derivative of the transfer matrix a hamiltonian with
terms that couple every pair of sites in the bulk with  the
boundary. That is, a  non local hamiltonian arises.\\
The hamiltonians associated to the first family of solutions
(\ref{sola}) are given by:

\begin{eqnarray}
H=\sum_{i=1}^{N-1}h^{XYZ}_{n,n+1}+\xi_{-}\sigma^{\alpha}_{1}+
\xi_{+}\sigma^{\beta}_{N}
\label{hyp}
\end{eqnarray}

Here $\alpha$ and $\beta$ can take the values $x$ or $y$ in all
possible combinations and the $\xi_{\pm}$ are arbitrary parameters
proportional to $\lambda_{HA}$.\\
As is clear, by rotating the axis, we can make the indices $\alpha$
and $\beta$ in equation (\ref{hyp}) take also the value z.\\
Equation (\ref{hyp}) gives the most general choice of boundary
conditions compatible with integrability for the XYZ chain besides
periodic and twisted boundary conditions. By twisted boundary
conditions we mean:

\begin{equation}
\sigma^{\alpha}_{N+1}=M\sigma^{\alpha}_{1}M^{-1}\;\;\;.
\end{equation}

Where $\alpha=x,y,z$ and the twisting matrix $M$ stands for a discrete
symmetry of the eight-vertex model. That is, $M=\sigma^{z}$ or $\sigma^{x}$.\\
In conclusion, the XYZ hamiltonian is integrable with boundary
conditions that correspond to the coupling with a magnetic
 field on the end sites
 oriented along parallel or orthogonal directions.
\end{section}

\begin{section}{General K-matrices for the Six Vertex model
(XXZ and XXX chains)}
In this section, we briefly review the results of \cite{nos} concerning
the general solution for the K-matrices of the XXZ chain and give the
general solution for the XXX case.\\
The R-matrix of the six vertex model is given by:

\begin{equation}
R(\theta)=\left(\begin{array}{cccc}
1 & 0 & 0 & 0\\
0 & \frac{\sin{\gamma}}{\sin{(\theta+\gamma)}} &
\frac{\sin{\theta}}{\sin{(\theta+\gamma)}} & 0\\
 0 & \frac{\sin{\theta}}  {\sin{(\theta+\gamma)}} &
\frac{\sin{\gamma}}  {\sin{(\theta+\gamma)}}  & 0\\
0 & 0 & 0 & 1
\end{array}\right)
\;\;\;,\label{Rt}
\end{equation}

and the general solution to the factorization equations in this model
is given by \cite{nos}:

\begin{equation}
K(\theta,\beta,\lambda,\mu,\xi)=\left( \begin{array}{cc}
\beta\sin(\xi+\theta) & \mu \sin2\theta \\
\lambda \sin2\theta & \beta\sin(\xi - \theta)
\end{array}  \right)
\;\;\;,\label{KG}
\end{equation}

 where $\beta,\xi,\mu$ and $\lambda$ are arbitrary parameters. The
associated hamiltonians to this K-matrix follow by the procedure
used above. Defining

\begin{equation}
K^{\pm}(\theta)=K(\theta,\beta_{\pm},\lambda_{\pm},\mu_{\pm},\xi_{\pm})
\;\;\;,\label{Kpm}
\end{equation}

the following hamiltonians are obtained:

\begin{eqnarray}
H =
\sum_{n=1}^{N-1}\left(\sigma^{x}_{n}\sigma^{x}_{n+1}+
\sigma^{y}_{n}\sigma^{y}_{n+1}+
\cosh\gamma \; \sigma^{z}_{n}\sigma^{z}_{n+1}\right)
\nonumber \\
+\sin\gamma\left(b_{-}\sigma_{1}^{z}-b_{+}\sigma_{N}^{z}+
c_{-}\sigma_{1}^{-}-c_{+}\sigma_{N}^{-}+
d_{-}\sigma_{1}^{+}-d_{+}\sigma_{N}^{+}\right)
\;\;\;,\label{Hpm}
\end{eqnarray}

where the parameters $b_{\pm}$, $c_{\pm}$ and $d_{\pm}$ follow
from $\lambda_{\pm}$, $\mu_{\pm}$, $\xi_{\pm}$ and $\beta_{\pm}$ as shown:

\begin{eqnarray}
b_{\pm}&=&\cot\xi_{\pm}\nonumber\\
c_{\pm}&=&\frac{2\lambda_{\pm}}{\beta_{\pm}\sin\xi_{\pm}}\nonumber\\
d_{\pm}&=&\frac{2\mu_{\pm}}{\beta_{\pm}\sin\xi_{\pm}}
\;\;\;.\label{cam}
\end{eqnarray}

Here $\beta_{\pm},\xi_{\pm}\neq 0$ so as to have $det[K^{-}(0)]\neq 0$ and
$tr[K^{+}(-\eta)]\neq 0$.\\
Equation (\ref{Hpm}) gives the most general choice of boundary
terms compatible with integrability for the XXZ chain besides periodic
and twisted b.c. In the present case one can twist the boundary
conditions as:

\begin{equation}
\sigma^{\alpha}_{N+1}=M\sigma^{\alpha}_{1}M^{-1}\;\;\;,
\end{equation}

where $M=\sigma^{x}$ or $M=e^{i\omega\sigma_{z}}$, $0<\omega<2\pi$.\\
Looking to equations (\ref{2}) and (\ref{4})
is now  possible to see
why in the elliptic case we lose a continuous parameter that appears in
the trigonometric limit. In the trigonometric case $\dpl=\dm=0$ and we
have only the constraint of equation (\ref{4}) which gives a continuous
family of solutions. The same happens with eqs. (\ref{1}) and
(\ref{3}) losing again a continuous parameter from the elliptic case.\\
It is also interesting to see what hamiltonians are obtained from the
trigonometric limit of the K-matrices obtained in the preceding
section. When  $k=0$ in (\ref{sola}) one obtains:

\begin{equation}
K_{TA}(\theta)=\left( \begin{array}{cc}
1& \epsilon\lambda_{HA}\sin\theta \\
\lambda_{HA} \sinh\theta & 1
\end{array}  \right)
\;\;\;,\label{lka}
\end{equation}

and this is seen to correspond to solution (\ref{KG}) whith $\beta=1$,
 $\xi=\pm\frac{\pi}{2}$ and $\mu,\lambda=\pm\lambda_{HA}/2$. The
correspondig
 hamiltonians are
obtained from the substitution of these values of the parameters in
eqs.
 (\ref{Kpm}),
(\ref{Hpm}) and  (\ref{cam}).\\
For solution (\ref{solb}) when $k=0$ one obtains:

\begin{equation}
K_{TB}(\theta)=\left( \begin{array}{cc}
1& \epsilon\lambda_{HB}\cos\theta \\
\lambda_{HB} \cos\theta & -1
\end{array}  \right)
\;\;\;,\label{lkb}
\end{equation}

that coresponds to  solution (\ref{KG}) with $\beta=1$, $\xi=0,\pi$
and  $\mu,\lambda=\pm\lambda_{HB}/2$. As discussed
in the previous section this limit  leads to a hamiltonian
 which includes a non-local coupling with the boundaries.\\
 It is interesting to note at this point that the
trigonometric limit of the K-matrices for the 8-vertex model does not
lead to an $SU_{q}(2)$ invariant hamiltonian. This is not the case
for the free fermion 8-vertex model where the $CH_{q}(2)$ symmetry
given by the elliptic K-matrices ``contracts'' to a $U_{q}(gl(1,1))$
 symmetry in the trigonometric limit \cite{royo},\cite{rogo}.\\
Let us now look for the general solution of the factorization equations
in the rational limit of the R-matrix (\ref{Rt}) given by:

\begin{equation}
R(\theta)=\left(\begin{array}{cccc}
1 & 0 & 0 & 0\\
0 & \frac{1}{(\theta+1)} &
\frac{\theta}{(\theta+1)} & 0\\
 0 & \frac{\theta}{(\theta+1)} &
\frac{1}{(\theta+1)}  & 0\\
0 & 0 & 0 & 1
\end{array}\right)
\;\;\;.\label{Rr}
\end{equation}

The equations for the K-matrix when $R$ is rational just follow
by   substituting the sine functions by their arguments
(that is, $\sin(\omega)$ by $\omega$) in all the equations.
The number of independent equations is the same in the rational
and trigonometric cases. (This number decreases going from the
elliptic to the  trigonometric case). Thus, the general solution is:

\begin{equation}
K(\theta,\beta,\lambda,\mu,\xi)=\left( \begin{array}{cc}
\beta(\xi+\theta) & \mu \theta \\
\lambda \theta & \beta(\xi - \theta)
\end{array}  \right)
\;\;\;,\label{Kr}
\end{equation}

and using equation (\ref{H}) one obtains:

\begin{eqnarray}
H =
\sum_{n=1}^{N-1}\left(\sigma^{x}_{n}\sigma^{x}_{n+1}+
\sigma^{y}_{n}\sigma^{y}_{n+1}+\sigma^{z}_{n}\sigma^{z}_{n+1}\right)
\nonumber \\
+b_{-}\sigma_{1}^{z}-b_{+}\sigma_{N}^{z}+
c_{-}\sigma_{1}^{-}-c_{+}\sigma_{N}^{-}+
d_{-}\sigma_{1}^{+}-d_{+}\sigma_{N}^{+}
\;\;\;,\label{Hrpm}
\end{eqnarray}

where we have scaled by a factor of $2\gamma$ and omitted a term
proportional to the identity operator. In this case the parameters $b_{\pm}$,
$c_{\pm}$ and $d_{\pm}$ follow
from $\lambda_{\pm}$, $\mu_{\pm}$, $\xi_{\pm}$ and $\beta_{\pm}$ as:

\begin{eqnarray}
b_{\pm}&=&\frac{1}{\xi_{\pm}}\nonumber\\
c_{\pm}&=&\frac{\lambda_{\pm}}{\beta_{\pm}\xi_{\pm}}\nonumber\\
d_{\pm}&=&\frac{\mu_{\pm}}{\beta_{\pm}\xi_{\pm}}
\;\;\;,\label{cam2}
\end{eqnarray}

where $\beta_{\pm},\xi_{\pm}\neq 0$ to have $det[K^{-}(0)]\neq 0$ and
$tr[K^{+}(-\eta)]\neq 0$.\\
This equation again provides the most general choice of boundary
terms compatible with integrability for the XXX chain besides periodic
and twisted b.c.

\section{Conclusions}

We have presented the general solution to the surface factorization equations
for the XYZ, XXZ and XXX models providing in this way the most general
boundary terms compatible with integrability. One can expect that if
any kind of quantum group invariance is possible in the XYZ chain the
 necessary boundary terms will be provided by those of
 hamiltonian (\ref{hyp}). For the XYZ chain a
geneneralization of the construction for the eigenvalues and
eigenvectors of the periodic chain remains to be done. As the
hamiltonians obtained for
the XXZ and XXX models do not commute with $J_{z}$ a  generalization of the
Functional  Bethe ansatz proposed by Sklyanin \cite{nan} for open boundary
conditions should be useful to find the eigenvalues.\\
In the context of two dimensional integrable quantum field theories
with boundaries it is interesting to solve the boundary bootstrap and
 cross-unitarity equations for these solutions.

\vspace{3cm}

A.G.R. would like to thank the LPTHE for the kind hospitality and the Spanish
M.E.C for financial support under grant AP90 02620085.
\end{section}

\end{document}